\documentclass[11pt, a4paper, twoside, notitlepage, openany]{article}

\usepackage[T1]{fontenc}
\usepackage{amsmath}
\usepackage{amssymb}
\usepackage{graphicx}
\usepackage{tabularx}
\usepackage{fullpage}
\usepackage{bm}

\newcommand{\absnum}{\ensuremath{\mathnormal{N}}}
\newcommand{\absnumrot}{\ensuremath{\absnum_\mathrm{DOF}^\mathrm{rot}}}
\newcommand{\absnumvec}{\ensuremath{\mathbf{N}}}

\newcommand{\adsorptionvec}{\ensuremath{\mathbf{\bm\Gamma}}}

\newcommand{\angmomk}{\ensuremath{\mathnormal{p_\natiter'}}}
\newcommand{\angmomvec}{\ensuremath{\mathbf{p'}}}
\newcommand{\axis}{\ensuremath{\mathnormal{\ell}}}
\newcommand{\axisb}{\ensuremath{\mathnormal{m}}}
\newcommand{\area}{\ensuremath{\mathnormal{a}}}
\newcommand{\boundary}{\ensuremath{\mathbb{B}}}
\newcommand{\bulklabel}{\ensuremath{\mathnormal{\beta}}}

\newcommand{\chempotvec}{\ensuremath{\mathbf{\bm\mu}}}
\newcommand{\component}{\ensuremath{\mathnormal{\kappa}}}
\newcommand{\compnum}{\ensuremath{\mathnormal{n}}}
\newcommand{\compvec}{\ensuremath{\mathbf{x}}}
\newcommand{\coord}{\ensuremath{\mathnormal{q}}}
\newcommand{\coordvec}{\ensuremath{\mathbf{q}}}
\newcommand{\coordr}{\ensuremath{\mathnormal{r}}}
\newcommand{\coordx}{\ensuremath{\mathnormal{x}}}
\newcommand{\coordy}{\ensuremath{\mathnormal{y}}}
\newcommand{\coordz}{\ensuremath{\mathnormal{z}}}
\newcommand{\density}{\ensuremath{\mathnormal{\rho}}}
\newcommand{\densityvec}{\ensuremath{\mathbf{\bm\rho}}}
\newcommand{\diff}{\ensuremath{\mathnormal{d}}}
\newcommand{\diffvec}{\ensuremath{\mathbf{d}}}
\newcommand{\dispersedlabel}{\ensuremath{\mathnormal{\alpha}}}
\newcommand{\distance}{\ensuremath{\mathnormal{r}}}
\newcommand{\distanceij}{\ensuremath{\distance_{\moli\molj}}}
\newcommand{\distancevec}{\ensuremath{\mathbf{r}}}
\newcommand{\distort}{\ensuremath{\mathnormal{\Lambda}}}
\newcommand{\distortten}{\ensuremath{\mathbf{\bm\Lambda}}}
\newcommand{\energy}{\ensuremath{\mathnormal{E}}}
\newcommand{\energyint}{\ensuremath{\mathnormal{U}}}
\newcommand{\energykin}{\ensuremath{\mathnormal{U}_\mathrm{kin}}}
\newcommand{\energykinrotk}{\ensuremath{\mathnormal{U}_\mathrm{kin, \natiter}^\mathrm{rot}}}
\newcommand{\energykintransk}{\ensuremath{\mathnormal{U}_\mathrm{kin, \natiter}^\mathrm{trans}}}
\newcommand{\energypot}{\ensuremath{\mathnormal{U}_\mathrm{pot}}}
\newcommand{\entropy}{\ensuremath{\mathnormal{S}}}
\newcommand{\equimolarlabel}{\ensuremath{\mathrm{e}}}
\newcommand{\excesslabel}{\ensuremath{\mathrm{E}}}
\newcommand{\extensive}{\ensuremath{\mathnormal{\Phi}}}
\newcommand{\force}{\ensuremath{\mathnormal{f}}}
\newcommand{\forceij}{\ensuremath{\force_{\moli\molj}}}
\newcommand{\forcevec}{\ensuremath{\mathbf{f}}}
\newcommand{\grandpot}{\ensuremath{\mathnormal{\Omega}}}
\newcommand{\hamilton}{\ensuremath{\mathcal{H}}}
\newcommand{\inertia}{\ensuremath{\mathnormal{I}}}
\newcommand{\helmholtz}{\ensuremath{\mathnormal{F}}}
\newcommand{\intensive}{\ensuremath{\mathnormal{\varphi}}}
\newcommand{\kinetic}{\ensuremath{\mathnormal{K}}}
\newcommand{\kineticsym}{\ensuremath{\mathnormal{p}}}
\newcommand{\kineticten}{\ensuremath{\mathbf{K}}}
\newcommand{\laplacelabel}{\ensuremath{\mathrm{L}}}
\newcommand{\laua}{\ensuremath{\mathnormal{\alpha}}}
\newcommand{\laub}{\ensuremath{\mathnormal{\beta}}}
\newcommand{\length}{\ensuremath{\mathnormal{L}}}
\newcommand{\mass}{\ensuremath{\mathnormal{m}}}
\newcommand{\moli}{\ensuremath{\mathnormal{i}}}
\newcommand{\molj}{\ensuremath{\mathnormal{j}}}
\newcommand{\momentum}{\ensuremath{\mathnormal{p}}}
\newcommand{\momentumvec}{\ensuremath{\mathbf{p}}}
\newcommand{\natiter}{\ensuremath{\mathnormal{k}}}
\newcommand{\naturals}{\ensuremath{\mathbb{N}}}
\newcommand{\normallabel}{\ensuremath{\mathrm{n}}}
\newcommand{\notion}{\ensuremath{\mathrm{\nu}}}
\newcommand{\orderof}{\ensuremath{\mathcal{O}}}

\newcommand{\orientvec}{\ensuremath{\mathbf{q'}}}
\newcommand{\orientveci}{\ensuremath{\mathbf{q_\moli'}}}
\newcommand{\orientvecj}{\ensuremath{\mathbf{q_\molj'}}}
\newcommand{\phasealabel}{\ensuremath{\mathnormal{A}}}
\newcommand{\phaseblabel}{\ensuremath{\mathnormal{B}}}
\newcommand{\potential}{\ensuremath{\mathnormal{u}}}
\newcommand{\potentialij}{\ensuremath{\potential_{\moli\molj}}}
\newcommand{\pressure}{\ensuremath{\mathnormal{P}}}
\newcommand{\pressurediffabsnu}{\ensuremath{\mathnormal{\Delta\pressure^\sigma_\notion}}}
\newcommand{\pressureten}{\ensuremath{\mathbf{P}}}
\newcommand{\spatialvec}{\ensuremath{\mathbf{r}}}
\newcommand{\surfentrop}{\ensuremath{\mathnormal{\zeta}}}
\newcommand{\radius}{\ensuremath{\mathnormal{R}}}
\newcommand{\radiusequim}{\ensuremath{\radius_\mathrm{e}}}
\newcommand{\tangentiallabel}{\ensuremath{\mathrm{t}}}
\newcommand{\temperature}{\ensuremath{\mathnormal{T}}}
\newcommand{\tensionabs}{\ensuremath{\mathnormal{\sigma}}}
\newcommand{\tensiondiff}{\ensuremath{\mathnormal{\gamma}}}
\newcommand{\tensionpart}{\ensuremath{\mathnormal{\hat\gamma}}}
\newcommand{\transf}{\ensuremath{\mathnormal{\chi}}}

\newcommand{\unitmatrix}{\ensuremath{\mathbf{1}}}
\newcommand{\vircoeff}{\ensuremath{\mathnormal{B}}}
\newcommand{\virial}{\ensuremath{\mathnormal{\Pi}}}
\newcommand{\virialsym}{\ensuremath{\mathnormal{\Xi}}}
\newcommand{\virialten}{\ensuremath{\mathbf{\bm\Pi}}}
\newcommand{\volume}{\ensuremath{\mathnormal{V}}}

\newcommand{\avg}[1]{\ensuremath{\left< {#1} \right>}}
\newcommand{\bulk}[1]{\ensuremath{{#1}^\bulklabel}}
\newcommand{\dispersed}[1]{\ensuremath{{#1}^\dispersedlabel}}

\newcommand{\excess}[1]{\ensuremath{{#1}^\excesslabel}}
\newcommand{\laplace}[1]{\ensuremath{{#1}_\laplacelabel}}
\newcommand{\normal}[1]{\ensuremath{{#1}_\normallabel}}
\newcommand{\nucform}[1]{\ensuremath{\Delta{#1}^\mathnormal{\star}}}
\newcommand{\phasea}[1]{\ensuremath{{#1}^\phasealabel}}
\newcommand{\phaseb}[1]{\ensuremath{{#1}^\phaseblabel}}
\newcommand{\tangential}[1]{\ensuremath{{#1}_\tangentiallabel}}
\newcommand{\liq}[1]{\ensuremath{{#1}'}}
\newcommand{\vap}[1]{\ensuremath{{#1}''}}

\newcommand{\qq}[1]{``{#1}''}

\newcommand{\cf}{\textit{cf.}}
\newcommand{\eg}{\textit{e.g.}}
\newcommand{\etal}{\textit{et al.}}
\newcommand{\etc}{\textit{etc.}}
\newcommand{\ie}{\textit{i.e.}}

\title{Open questions on defining and computing the vapour-liquid surface tension by virial and test transformation approaches}
\author{Martin Thomas Horsch\footnote{Norwegian University of Life Sciences, Material Theory and Informatics Group, P.O.~Box 5003, 1432 \AA{}s, Norway. E-mail: martin.thomas.horsch@nmbu.no.}}
\date{}

\begin{document}

\maketitle

\begin{abstract}
This work addresses four problems in defining and computing the surface tension of vapour-liquid interfaces: (1)~The apparent kinetic contribution to the surface tension, and what it is that makes it appear; (2)~the problem of defining the local virial, which for some purposes appears to be necessary; (3)~the disagreement between results for spherical interfaces when using different methods, specifically the virial and the test area method; (4)~how the surface tension values obtained from these methods can be related to the definition of the surface tension from thermodynamics -- since if they cannot, it is a meaningless exercise to compute them. At the end, all these problems remain unsolved: They are \qq{open questions.}
\end{abstract}

\section{Introduction}

\subsection{Interfacial properties: Open questions}

Molecular modelling and simulation is often applied to the thermodynamic properties of fluid systems. Most commonly, the underlying models are defined by pairwise additive force fields which describe the intermolecular interactions by effective pair potentials. While thermodynamic properties, including for fluid phase equilibria, can be obtained by molecular simulation of homogeneous systems, also heterogeneous systems can be simulated with the same intermolecular pair potentials. In this way, in particular, interfacial properties can be computed, the most important of which is the surface tension. While this is regularly done and can be considered an established practice, fundamental \textit{open questions} remain that threaten to undercut what has been claimed to be the foundation for commonly employed methods. This in turn puts much of the results from the literature into question. Here we will be considering a few such open questions with a focus on methods for computing the surface tension that are based on the virial.

The local virial tensor (and pressure tensor) is always an underspecified quantity, but in the planar case, this does not matter because the volume integral over it yields a unique, well-defined result both for the overall pressure tensor and the surface tension. When integrating over the (virial as a function of the) radius in polar coordinates, this property disappears, and the surface tension is no longer well-defined.
Unfortunately, results for the surface tension of droplets and bubbles from the virial route are known to disagree both with results from test transformations (test area method) and with what is known about the Tolman length for the respective systems. In particular, the surface tension of
Lennard-Jones nanodroplets is significantly smaller if computed by the virial route than by the test area method~\cite{SMMMJ10}; other thermodynamic methods tend to agree with the test area results~\cite{BDOVB10}. Such a disagreement was found in the case of curved interfaces only, whereas for planar interfaces, the two approaches are exactly equivalent~\cite{SM91, GM12}. Lau \etal~\cite{LFHMJ15} uncovered a possible source of this disagreement by proving that for spherical interfaces, the response of the interfacial free energy to a small distortion (test transformation) contains a non-zero contribution from the variance of the response of the energy, \ie, from a second-order contribution in terms of the internal energy of the system.

As usual in molecular simulation, these methods are based on statistical mechanical expressions that are connected to phenomenological thermodynamics through the partition function and the thermodynamic potential. In the spherical case, however, this connection looks more shaky that it usually does when connecting thermodynamics to statistical mechanics. Moreover, even in the planar case, Sega \etal~\cite{SFJ17} argue that a kinetic contribution appears, and that this contribution would need to be added to that from the virial, which to date has not become common practice at all. Is this an indication that almost all molecular simulation data on the surface tension are wrong? In the case of water, for instance, the error due to neglecting the effect from the apparent kinetic contribution would amount to 15\%, Sega \etal~\cite{SFJ17} find.

These problems are summarized below; none of them will be resolved, instead, the aim here is to highlight their importance and provide some input to further discussions. This is proposed in the hope that the discussion also represents some actual progress on the issues mentioned. The apparent kinetic contribution to the surface tension of planar interfaces is discussed in Section~\ref{sec:planar}. Then the spherical case is discussed: The methodology and the underlying statistical mechanics in Section~\ref{sec:statmech}, the perspective from phenomenological thermodynamics in Section~\ref{sec:phenomenological}. However, after the first thermodynamic analysis, some problems remain. Concluding this work, an orientation for future efforts is proposed in Section~\ref{sec:conclusion}. This will all be preceded by a clarification of the scope of this work and the terminology to be used (Section~\ref{subsec:terminology}).

\subsection{Key concepts and scope of this work}
\label{subsec:terminology}

This section is meant as a terminological clarification. Key quantities discussed in this work are the \textit{virial} and the \textit{surface tension}. A term that may cause misunderstandings is \qq{second-order virial,} a concept introduced as central to the methodology presented in Section~\ref{sec:statmech}.

The word \textit{virial} is taken from Latin and means \textit{related to forces}. In thermodynamics, it is used for the \textit{virial expansion} (which this work is not about), the \textit{virial coefficients} which are the coefficients that occur in the virial expansion (this work is therefore not about these either), the quantity called \textit{the virial}, and its representation as a tensorial quantity, the \textit{virial tensor} (the latter two \textit{are} what this work is about). These all have in common that they describe how the behaviour of a fluid deviates from that of an ideal gas \textit{due to the forces} acting between molecules. The virial expansion, also called the virial equation of state, is a caloric equation of state
\begin{equation}
   \frac{\pressure}{\temperature} = \density + \vircoeff_2(\temperature)\density^2 + \vircoeff_3(\temperature)\density^3 + \vircoeff_4(\temperature)\density^4 + \dots,
\label{eqn:virial-expansion}
\end{equation}
when expanded in terms of density, where $\vircoeff_\natiter$ is called the $\natiter$-th virial coefficient. Thereof, the most important one is $\vircoeff_2$, the \textit{second virial coefficient}, which controls the leading contribution to the deviation between ideal and real fluid behaviour. If Eq.~(\ref{eqn:virial-expansion}) is truncated after the second term, $\pressure/\temperature = \density + \vircoeff_2\density^2$, it becomes an expansion to the second order, which is consequently referred to as a \textit{second-order virial expansion} or \textit{second-order virial equation of state}. Both of these terms contain the phrase \qq{second-order virial,} and they indeed have some relevance to the theory of nucleation and the surface tension of nanodroplets. For example, in Laaksonen-Ford-Kulmala nucleation theory~\cite{LFK94}, the free energy of formation for clusters in a supersaturated vapour is determined from the second-order virial expansion; the key parameter required for this is $\vircoeff_2$. The free energy of formation of the clusters can be converted to an estimate for their surface tension~\cite{HVBGRWSH08}. Urrutia~\cite{Urrutia14, Urrutia16} discusses the curvature dependence of the surface tension in terms of an expansion for the grand potential based on Mayer cluster integrals~\cite{MM40}, which can be truncated after its second term; for this, he uses the term \qq{second-order virial series,} including in the title of one of his articles~\cite{Urrutia16}. It is therefore appropriate to point out that here we will not be discussing these theories, and that our use of the term \textit{second-order virial} will refer to a generalization of the virial and the virial tensor, not the virial coefficient or the virial expansion.

The virial tensor, which has the dimension of an energy, is defined by
\begin{eqnarray}
   \virial^{\axis\axisb}(\coordvec, \orientvec) ~=~ \sum_{\{\moli, \molj\}} \distanceij^\axis\forceij^\axisb
     ~=~ - \sum_{\{\moli, \molj\}} \distanceij^\axis \frac{\partial\potentialij}{\partial\distanceij^\axisb},
\end{eqnarray}
denoted more compactly \eg~by $\virialten = \sum \distancevec_{\moli\molj}\otimes\forcevec_{\moli\molj}$,
for a system of rigid units interacting through the pair potential $\potentialij(\distancevec_{\moli\molj}, \orientveci, \orientvecj)$. In general, the pair potential depends on the species of $\moli$ and $\molj$. The configuration of the system is given by $(\coordvec, \orientvec)$; the first element contains the translational, the second one the rotational coordinates of all rigid units. The $\distancevec_{\moli\molj} = \coordvec_\molj - \coordvec_\moli$ are distance vectors between the rigid-unit centres of mass $\coordvec_\moli$ and $\coordvec_\molj$, while the $\orientveci$ and $\orientvecj$ represent any orientation coordinates of the two molecules $\moli$ and $\molj$, corresponding to any rotational degrees of freedom, if present.
Superscript $\axis$ and $\axisb$ symbols are indices representing coordinate axes.

To enable the common notational conventions of the type $\force = -\nabla\potential$, the signs for the quantities need to be defined appopriately. So if $\distance_{\moli\molj}^\axis = \coord_\molj^\axis - \coord_\moli^\axis$ denotes the distance from $\moli$ to $\molj$ in $\axis$ direction, a positive force in that direction must tend to increase that value. Accordingly, $\forcevec_{\moli\molj}$ is here defined to be the force acting on $\molj$ due to the interaction between $\moli$ and $\molj$. The force acting on $\moli$ from this interaction will have the same magnitude and the opposite sign.

The virial tensor $\virialten$, a statistical-mechanical observable, is related to the pressure tensor
\begin{equation}
   \pressureten = \unitmatrix\density\temperature + \frac{\avg{\virialten}}{\volume},
\label{eqn:standard-pressure-tensor}
\end{equation}
by ensemble averaging in the canonical ensemble (and very similarly in other ensembles), where $\unitmatrix$ is the unit matrix.
The scalar quantities corresponding to $\virialten$ and $\pressureten$ are the virial
\begin{equation}
   \virial(\coordvec, \orientvec) ~=~ \mathrm{tr} \, \virialten ~=~ \virial^{\coordx\coordx} + \virial^{\coordy\coordy} + \virial^{\coordz\coordz}
      ~=~ \sum_{\{\moli, \molj\}} \distancevec_{\moli\molj} \forcevec_{\moli\molj}
      ~=~ - \sum_{\{\moli, \molj\}} \distancevec_{\moli\molj} \nabla_{\distancevec_{\moli\molj}} \potentialij,
\end{equation}
and the pressure
\begin{equation}
   \pressure ~=~ - \left(\frac{\partial\helmholtz}{\partial\volume}\right)_{\absnumvec, \temperature}
      ~=~ \frac{1}{3} \, \mathrm{tr} \, \pressureten
      ~=~ \density\temperature + \frac{\avg\virial}{3\volume}.
\end{equation}
The virial $\virial$ characterizes the residual of $\pressure$, which is a first derivative of Helmholtz free energy~$\helmholtz$; also, $\virial$ is a microscopic observable determined by summation over first derivatives of the pair potential. Extending this by analogy, in Section~\ref{sec:statmech}, higher-order virials will be introduced, where the $k$-th order virial tensor contains the $k$-th derivatives of the pair potential.

\begin{figure}[b!]
    \centering
    \includegraphics[width=0.5\textwidth]{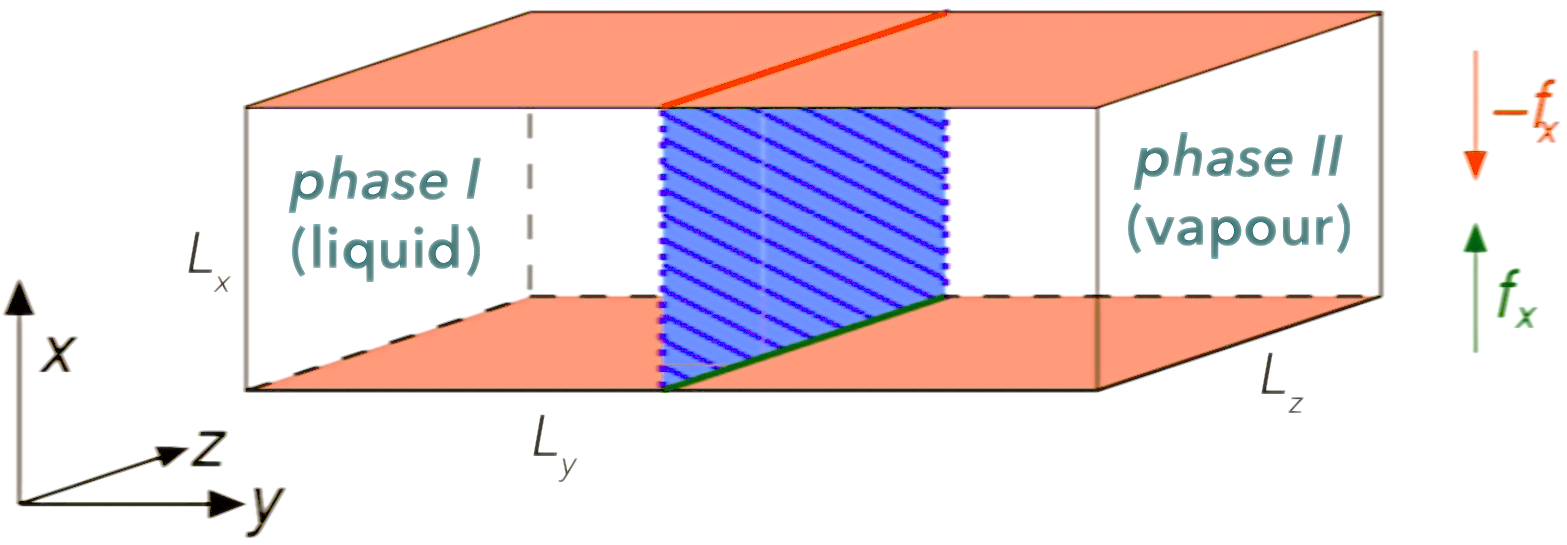}
    \caption{Macromechanics of surface tension for a planar interface with orientation normal to the $\coordy$ axis. The surface tension is the contracting force per contact line length, $\tensiondiff = \force_\coordx / \length_\coordz$.}
    \label{fig:planar-geometry}
\end{figure}

The \textit{surface tension}, from the point of view of \textit{macroscopic mechanics}, is the work (per surface area) required to increase the area of a phase boundary, or the force (per contact line length) that the interface would apply to a third body that it is in contact with, such as a wall, expressing its tendency to contract itself in order to reduce its area.

For a system or control volume with a geometry as shown in Fig.~\ref{fig:planar-geometry}, the two coexisting fluid phases are at mechanical equilibrium with the same pressure, $\liq\pressure = \vap\pressure = \pressure_\mathrm{bulk}$, since the interface is not curved. The pressure in normal direction ($\coordy$ axis) is simply given by this value $\normal\pressure = \pressure^{\coordy\coordy} = \pressure_\mathrm{bulk}$. In tangential direction ($\coordx$ and $\coordz$ axes), the surface tension acts as a contracting force proportional to the length of its contact line with the respective control volume boundary. As highlighted in Fig.~\ref{fig:planar-geometry}, the face normal to the $\coordx$ axis experiences the contracting force $\force_\coordx = \tensiondiff\length_\coordz$, and since the area of that face is $\length_\coordy\length_\coordz$, this contributes $-\force_\coordx/\length_\coordy\length_\coordz = -\tensiondiff/\length_\coordy$ to pressure in $\coordx$ direction, which becomes $\pressure^{\coordx\coordx} = \pressure_\mathrm{bulk} - \tensiondiff/\length_\coordy$. In $\coordz$ direction, the contracting force is $\force_\coordz = \tensiondiff\length_\coordx$, the box boundary area is $\length_\coordx\length_\coordy$, and the resulting pressure is $\pressure^{\coordz\coordz} = \pressure_\mathrm{bulk} - \tensiondiff/\length_\coordy$, same as in $\coordx$ direction; this is the tangential pressure, $\tangential\pressure = \pressure^{\coordx\coordx} = \pressure^{\coordz\coordz}$. Note that the surface area of the vapour-liquid interface in this example is $\area = \length_\coordx\length_\coordz = \volume/\length_\coordy$. We can therefore write
\begin{eqnarray}
   \tensiondiff &=& (\normal\pressure - \tangential\pressure) \, \length_\coordy, \label{eqn:macromech-pn-pt-dy} \\
   \tensiondiff &=& (\normal\pressure - \tangential\pressure) \, \volume\area^{-1},
\end{eqnarray}
for the case of a single planar interface oriented normally to the $\coordy$ axis.

While this applies to a macroscopic system, we can also directly evaluate it for a molecular simulation box with an analogous geometry. Normally, then, there are two such interfaces, so that the periodic boundary condition can be satisfied.\footnote{In that case, $(\normal\pressure - \tangential\pressure) \, \length_\coordy$ evaluates to $2\tensiondiff$, because there are two interfaces that contribute. Their total surface area is $\area = 2\length_\coordx\length_\coordz$, and accounting for this, $\tensiondiff = (\normal\pressure - \tangential\pressure)\volume\area^{-1}$ still holds.} Here, for simplicity, let us continue to consider the control volume with a single vapour-liquid interface. The \textit{virial route} to the surface tension consists in converting this to an expression in terms of the virial
\begin{eqnarray}
   \normal\pressure & = & \density\temperature \,+\, \frac{\avg{\virial^{\coordy\coordy}}}{\volume}, \\
   \tangential\pressure & = & \density\temperature \,+\, \frac{\avg{\virial^{\coordx\coordx}} + \avg{\virial^{\coordz\coordz}}}{2\volume}, \label{eqn:pressure-xz} \\
   \tensiondiff & = & \frac{1}{\area} \, \left(\avg{\virial^{\coordy\coordy}} - \frac{\avg{\virial^{\coordx\coordx}} + \avg{\virial^{\coordz\coordz}}}{2}\right),
\label{eqn:virialroute-macro-planar}
\end{eqnarray}
where the averaging over $\coordx$ and $\coordz$ in Eqs.~(\ref{eqn:pressure-xz}) and (\ref{eqn:virialroute-macro-planar}) is only done in order to improve sampling.

This virial route can be implemented in a molecular simulation code with little effort; only a single scalar observable needs to be sampled, namely, $(\virial^{\coordy\coordy} - [\virial^{\coordx\coordx} + \virial^{\coordz\coordz}]/2)(\coordvec, \orientvec)$, which for any given configuration $(\coordvec, \orientvec)$ expresses the anisotropy of the virial tensor. However, in addition to this macromechanically motivated perspective, the virial route can also be implemented micromechanically in terms of a localized pressure tensor $\pressureten(\spatialvec)$; usually, the symmetry of the system is taken into account and only profiles over the normal coordinate are sampled, so that this becomes $\pressureten(\coordy)$. Then, Eq.~(\ref{eqn:macromech-pn-pt-dy}) can be evaluated by integration over the $\coordy$ axis\footnote{Again, this is true if the system contains \textit{one} interface normal to $\coordy$, like in Fig.~\ref{fig:planar-geometry}. If there are two such interfaces, the integral over the whole box will yield $2\tensiondiff$.}
\begin{equation}
   \tensiondiff = \int_0^{\length_\coordy} \left[\normal\pressure(\coordy) - \tangential\pressure(\coordy)\right] \, \diff\coordy.
\label{eqn:virialroute-micro-planar}
\end{equation}
Both the ideal and the residual pressure will vary as a function of $\coordy$, so that for computing $\normal\pressure(\coordy)$ and $\tangential\pressure(\coordy)$, the density profile needs to be sampled in addition to the profiles of the normal and tangential virial. When taking the difference between both, of course, the ideal part cancels out again, so it is enough to take an integral over the anisotropy of the local virial. Other than the virial-route method from Eq.~(\ref{eqn:virialroute-micro-planar}) requiring more data than the one from Eq.~(\ref{eqn:virialroute-macro-planar}), these two descriptions of the virial route are just variants of each other -- at least for planar interfaces, they are strictly equivalent and can also be deduced directly from the canonical partition function. So at least the planar surface tension is well understood, we would think based on the above.

Sega \etal~\cite{SFJ17, SJ18, SHJ18}, however, would disagree. From their point of view, we made a mistake already in Eq.~(\ref{eqn:standard-pressure-tensor}), where we failed to properly account for the contribution of molecular motion to the pressure tensor, the complete definition of which contains the kinetic energy tensor $\kineticten$,
\begin{equation}
   \pressureten ~=~ \frac{\avg{\kineticten} + \avg{\virialten}}{\volume}
      ~=~ \frac{1}{\volume} \, \left<
         \sum_\moli \mass_\moli^{-1}\,\momentumvec_\moli \otimes \momentumvec_\moli
            \, + \, 
         \sum_{\{\moli, \molj\}} \distancevec_{\moli\molj} \otimes \forcevec_{\moli\molj}
      \right>.
\label{eqn:kinetic-pressure-tensor}
\end{equation}
If this is accounted for consistently, they argue, a kinetic contribution appears to be present in $\tensiondiff$ whenever $\kineticten$ is anisotropic, and in the case of molecules represented as rigid units that are not radially symmetric, but have rotational degrees of freedom, it \textit{can} be anisotropic. Namely, that is the case whenever the molecules that can rotate have a \textit{preferential orientation} at the interface (or anywhere else in the system). We will continue to address this argument in Section~\ref{sec:planar}.

\section{Planar interfaces and the apparent kinetic contribution}
\label{sec:planar}

\begin{figure}[b!]
    \centering
    \includegraphics[width=0.9333\textwidth]{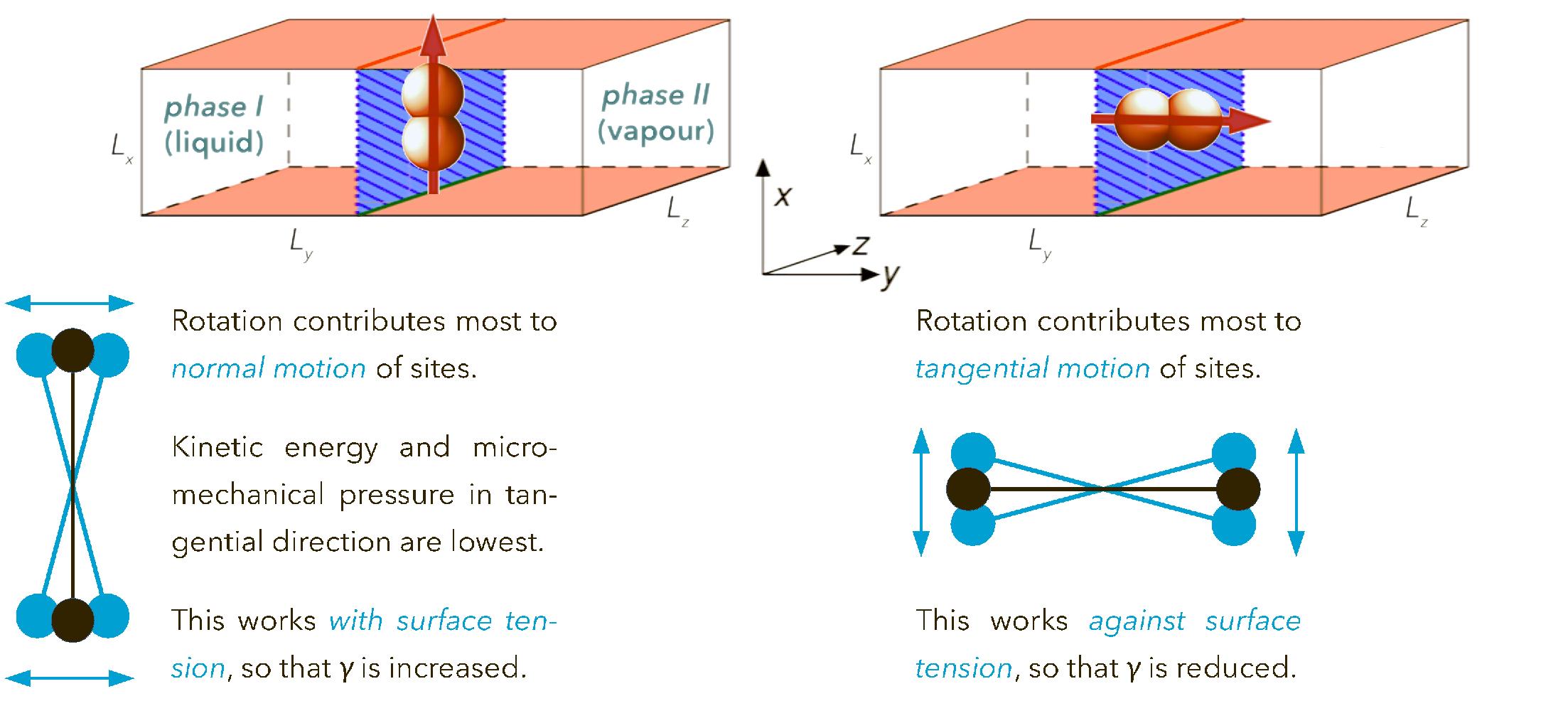}
    \caption{Intuition behind the apparent kinetic contribution illustrated for the case of an elongated molecular model. Where there is a preferential orientation, the motion of the interaction sites at the ends of the axis is not Maxwell distributed. Instead, where an orientation of the axis in tangential direction is preferred, the contribution of these sites to the kinetic energy in normal direction, $\kinetic^{\coordy\coordy}$, will be greater than to the kinetic energy in tangential direction, $\kinetic^{\coordx\coordx}$ and $\kinetic^{\coordz\coordz}$. In normally oriented molecules, the sites move more in tangential than in normal direction.}
    \label{fig:rotation}
\end{figure}

\subsection{The argument for the apparent kinetic contribution}

The apparent kinetic contribution (\qq{ideal gas contribution}) to the surface tension of planar interfaces was proposed by Sega \etal~\cite{SFJ17} for use in a particular frame of reference: Rigid multi-site models using \textit{individual interaction-site mechanics with constraints}. Many molecular simulation solvers, including \textit{ls1 mardyn}~\cite{NB3EHWBGHVH14} and \textit{ms2}~\cite{FGNSMLPCBSKHV21}, evaluate rigid multi-site models, instead, by \textit{mechanics over the rigid-unit centre of mass}. Does ls1 need to include the apparent kinetic energy contribution when computing $\tensiondiff$? And are all the previous results on the surface tension of multi-site models using industry-standard molecular simulation solvers incorrect?

The intuitive argument for the apparent kinetic contribution, as developed by Sega and Jedlovszky~\cite{SJ18}, is illustrated in Fig.~\ref{fig:rotation}. In the formalism based on individual-site mechanics with constraints due to rigid bonds, the position and momentum coordinates are not independent and, hence, cannot be separated in the partition function. The momenta are not Maxwell distributed; instead, a preferential normal orientation induces a preferential tangential motion, and vice versa. Therefore, the pressure tensors from Eq.~(\ref{eqn:standard-pressure-tensor}) and~(\ref{eqn:kinetic-pressure-tensor}) are not equal either. To distinguish the two, we here denote the virial-based statistical-mechanical pressure tensor from Eq.~(\ref{eqn:standard-pressure-tensor}) by $\pressureten^\virialsym$ and the micromechanical kinetic pressure tensor from Eq.~(\ref{eqn:kinetic-pressure-tensor}) by $\pressureten^\kineticsym$. Taking the anisotropy expression from Eq.~(\ref{eqn:virialroute-macro-planar}) leads to different results for the surface tension, which Sega \etal~\cite{SFJ17} refer to as $\tensiondiff^\virialsym$ and $\tensiondiff^\kineticsym$, respectively; out of these, they assert that $\tensiondiff^\kineticsym$ is the correct value.


\subsection{The argument against the apparent kinetic contribution}


In contrast to the above, I hold that these results and claims are the artefact of a method based on individual-site mechanics with a superimposed correction scheme for rigid bond lengths. With these methods, \ie, SHAKE~\cite{RCB77}, RATTLE~\cite{Andersen83}, or variants thereof~\cite{FS98, Gonnet07}, the equations of motion are complemented by additional projection steps: Sites first move according to the forces that they experience, but once they threaten to leave their orbit, an external entity intervenes and pushes them back (same idea as proposed by Newton, see Section~\ref{sec:conclusion}). But when, instead, representing a system of rigid units by centre-of-mass coordinates for translation (momenta $\momentumvec$ and positions $\coordvec$) and rotation (angular momenta $\angmomvec$ and orientations $\orientvec$), the Hamiltonian is
\begin{eqnarray}
   \hamilton(\momentumvec, \angmomvec, \coordvec, \orientvec) & = & \sum_{1 \leq \moli \leq \absnum} \frac{\momentum^2_{\moli\coordx} + \momentum^2_{\moli\coordy} + \momentum^2_{\moli\coordz}}{2\mass_\moli}
      + \sum_{1 \leq \natiter \leq \absnumrot} \frac{\angmomk^2}{2\inertia_\natiter}
      + \sum_{\{\moli, \molj\}} \potentialij(\distancevec_{\moli\molj}, \orientveci, \orientvecj) \nonumber \\
   & = & \sum_{1 \leq \natiter \leq 3\absnum} \energykintransk(\momentum_\natiter)
      ~+~ \sum_{1 \leq \natiter \leq \absnumrot} \energykinrotk(\angmomk)
      ~+~ \energypot(\coordvec, \orientvec).
\end{eqnarray}
Therein, $\absnumrot$ is the number of rotational degrees of freedom, and $\inertia_\natiter$ is the moment of inertia associated with the $\natiter$-th rotational degree of freedom. In other words, for the formalism used by ls1 mardyn~\cite{NB3EHWBGHVH14}, ms2~\cite{FGNSMLPCBSKHV21}, and similar codes, it is no problem at all not only to separate position and orientation coordinates, on the one hand, from the translational and angular momentum coordinates on the other hand. It is also, just as usual, possible to separate the contributions of all individual translational and rotational degrees of freedom from each other. Therefore, given that each of these degrees of freedom contributes to the kinetic energy through a quadratic term, each individual translational or angular momentum coordinate is Maxwell distributed.

In canonical simulations with an isokinetic thermostat, is not strictly true that we apply Hamiltonian dynamics: Momenta are rescaled by a factor very close to unity to maintain a constant total kinetic energy. Undesired effects from thermostat algorithms are known to create artefacts~\cite{HVR95}, \eg, in transport properties~\cite{HGVK22}. Here, there is no obvious reason to expect major problems; any error due to the thermostat should be of the order of its interventions. In the stationary state, after equilibration, these are minor, since they only compensate for the numerical error of the integrator. In the absence of such a numerical error, it would be possible to turn off the thermostat completely after equilibration and run a microcanonical simulation, and the temperature would remain constant, since constant $\absnumvec$, $\volume$, and $\energy$ at equilibrium determine $\temperature(\absnumvec, \volume, \energy)$. Alternatively, it is possible to use a Nos\'e-Hoover thermostat. So the small error due to the thermostat can be excluded in various ways, should it turn out to be an issue.


From the above, it would appear that the previous results on the surface tension of 2CLQ, 2CLD, and other multi-site molecules using the established method for planar interfaces, with $\pressureten = \pressureten^\virialsym$ and $\tensiondiff = \tensiondiff^\virialsym$, still stand; there, the code ls1 mardyn was used, and the equations of motion were evaluated over molecular centres of mass~\cite{WHH15a, WSKKHH15, WHH16, WHH17}. If this analysis is correct, it will not be necessary to implement the apparent kinetic contribution and repeat all the simulations.
A concise critique of the apparent kinetic contribution can be given as follows:
\begin{enumerate}
   \item{} In systems where molecules have a preferential orientation, the momentum of individual sites in a rigid multi-site model is not Maxwell distributed. Sega \etal~\cite{SFJ17, SJ18} and Lbadaoui-Darvas \etal~\cite{LIJ22} are making an important point when they advise not to ignore this. Consequently, over the coordinates of interaction sites, it is not possible to separate the partition function into a product of a momentum integral and a configurational integral. But is this not in fact a strong argument \textit{not to use interaction-site coordinates}, but rather centre-of-mass coordinates, when applying statistical mechanics to such a system?
   \item{} Mechanics over the translation and rotation of rigid-body centres of mass are a viable formalism for molecular dynamics simulation of rigid multi-site models. There, the kinetic energy and the potential energy can be separated in the Hamiltonian, so that they contribute to the partition function through separate factors. The kinetic factor does not depend on the volume of the system. Hence, when taking a derivative such as $\pressure^{\coordx\coordx} = -(\partial\diff\helmholtz/\partial\volume)_{\absnumvec, \length_\coordy, \length_\coordz, \temperature}$, the kinetic-energy contribution is constant and disappears when taking the derivative. Even when deducing the ideal gas law, the ideal term $\density\temperature$ is contributed by the integration boundary for the position coordinates in combination with the statistical prefactor; the momentum coordinates do not contribute to the pressure.
   \item{} However, even if the above was wrong and it was necessary to account for translational and angular momenta when computing $\pressureten$, their contribution would be isotropic. This is because the degrees of freedom can, over rigid-body centres of mass, all be separated from each other; therefore, each of them is Maxwell distributed, and there cannot be any preferential axis for which the contribution to $\kineticten$ is greater than for the others.
   \item{} It is plausible that for an integrator based on constraint mechanics, the apparent kinetic contribution must be included when computing $\tensiondiff$. The work by Sega~\etal~\cite{SFJ17} proves that there, $\tensiondiff^\virialsym$ and $\tensiondiff^\kineticsym$ are different, so they cannot both be correct. But that alone does not prove that $\tensiondiff^\virialsym$ is wrong and $\tensiondiff^\kineticsym$ is right. The argument that $\pressureten^\kineticsym$ is the pressure tensor with a thermodynamic meaning, and not $\pressureten^\virialsym$, would need to be developed from the partition function, which has not been done by Sega~\etal~\cite{SFJ17} or any of the follow-up works.
\end{enumerate}

\section{Spherical interfaces and the square perturbation contribution}
\label{sec:statmech}

The work by Lau \etal~\cite{LFHMJ15} on the test area method for spherical interfaces has established that, based on a definition of the surface tension as a Helmholtz free energy derivative in the canonical ensemble, both linear \textit{and quadratic} expressions in response to a test transformation need to be taken into account. Similarly, it has been established that test-area and virial-based methods disagree on the surface tension for spherical interfaces. It looks plausible that, similar to the test area method taking into account a second-order term for the response of the potential energy to a distortion, when considering a spherical geometry, a \textit{second-order virial} needs to be introduced.

\subsection{The virial route and non-uniqueness of the local virial}

\begin{figure}[b!]
    \centering
    \includegraphics[width=0.6\textwidth]{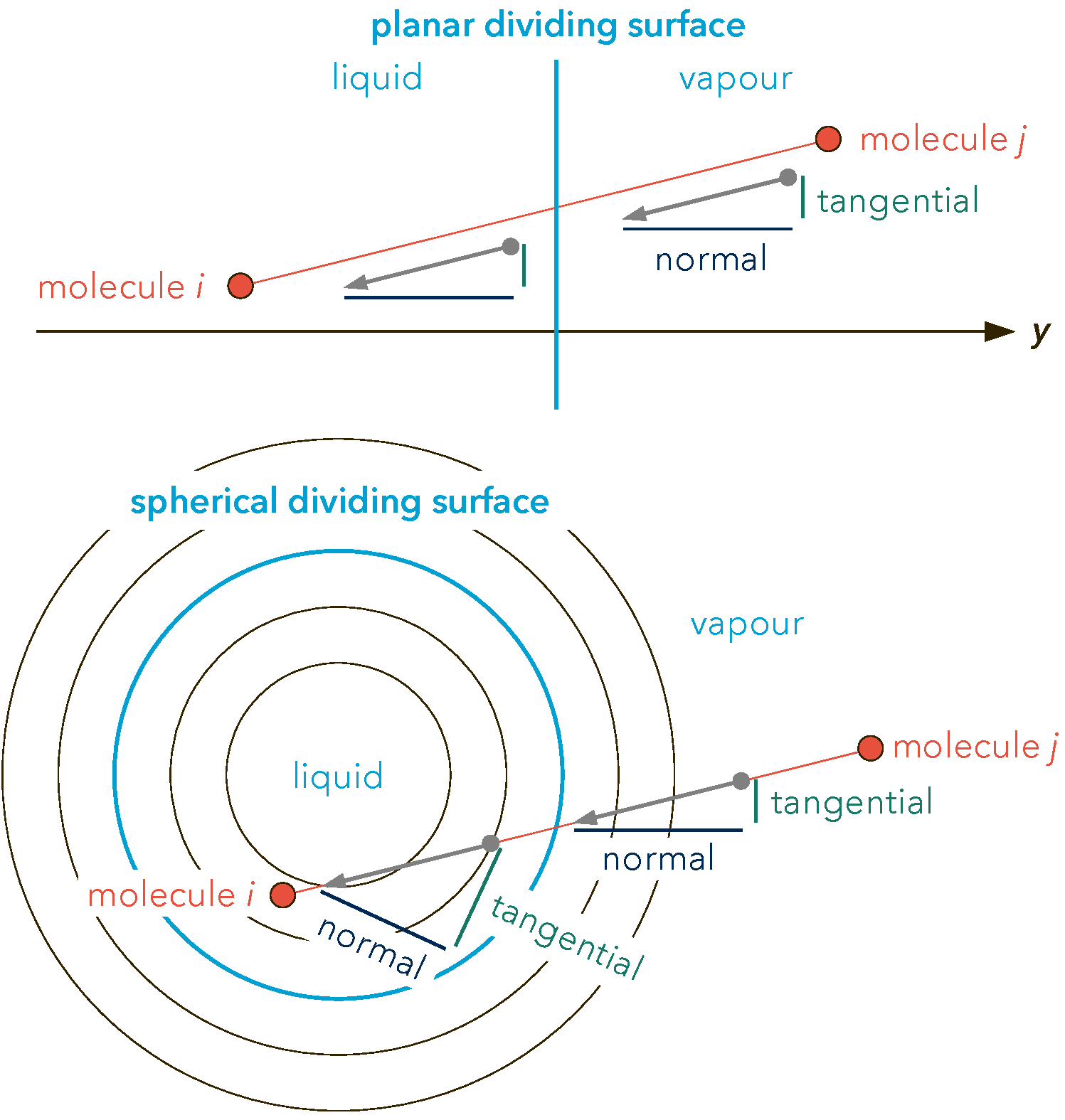}
    \caption{For a planar geometry, the interaction between molecules $\moli$ and $\molj$ contributes to the normal and the tangential pressure in a unique, well-defined way, irrespective of where that contribution is localized when constructing a profile over the normal coordinate. For a spherical geometry, that is not the case. The result from integration over the profiles depends on the scheme for localizing the virial, and it does not have a direct thermodynamic interpretation.}
    \label{fig:geometries}
\end{figure}

When evaluating the virial and the pressure in a localized way, a non-uniqueness emerges from the way the potential energy of the system is given: As a sum over pairwise intermolecular interactions. Two interacting molecules $\moli$ and $\molj$ each have a well-defined position, but the interaction between them does not. Depending on how the virial from their interaction is distributed over the coordinate over which the profile is taken, results for the local pressure profile can differ. Despite this, in the planar case, $\tensiondiff$ can be determined rigorously from the micromechanical integral over the normal and tangential pressure profiles as given in Eq.~(\ref{eqn:virialroute-micro-planar}) because the integral is unique, even though the integrand is not. The total always evaluates to the macromechanical expression from Eq.~(\ref{eqn:virialroute-macro-planar}), which is defined in terms of the total virial tensor (and total pressure tensor) of the system as a whole. In effect, the micromechanical virial route is nothing more than a visualization technique for the macromechanical virial route, useful in cases where it is desirable to show $\normal\pressure(\coordy)$ and $\tangential\pressure(\coordy)$ in diagrams in order to illustrate the results.

For spherical interfaces, that is not the case. To begin with, \textit{there is no macromechanical expression} that is independent of the scheme employed for localizing the contributions from each pairwise interaction. That issue will be addressed in this work; in Section~\ref{subsec:second-order-virial}, we will deduce such an expression based on the test-area method. In addition, different ways of distributing the virial from molecules $\moli$ and $\molj$ over space, \cf~Fig.~\ref{fig:geometries}, lead to different results for the integral over the profiles for the normal pressure $\normal\pressure(\coordr)$ and the tangential pressure $\tangential\pressure(\coordr)$ in radial coordinates
\begin{equation}
   \tensiondiff ~=~ \radius^{-2} \, \int_0^\infty \left(\normal\pressure(\coordr) - \tangential\pressure(\coordr)\right) \coordr^2 \diff\coordr,
\end{equation}
known as the Bakker-Buff equation~\cite{Bakker28, Buff55}; there, $\radius = \laplace\radius$ is the Laplace radius (\cf~Section~\ref{subsec:dividing-surface-notion}).

\subsection{Test area method for the surface tension of droplets and bubbles}
\label{subsec:spherical-test-area}

The basic idea behind the test area method goes back to Bennett~\cite{Bennett76} and, as regards the method in use today, to Gloor \etal~\cite{GJBM05}. For this purpose, consider two different two-phase canonical-ensemble systems with the same values of $\absnumvec$, $\volume$, and $\temperature$, but with different box shapes: In both cases, these are specified to be cuboid volumes with periodic boundary conditions, but one system has the dimensions $\length^0_\coordx \times \length^0_\coordy \times \length^0_\coordz = \volume$, while the other has $\length^1_\coordx \times \length^1_\coordy \times \length^1_\coordz = \volume$.

Applying phenomenological interfacial thermodynamics as underlying the method, all extensive and intensive properties of the two phases will be identical for both systems, including the volumes $\phasea\volume$ and $\phaseb\volume$, the compositions $\phasea\compvec$ and $\phaseb\compvec$, and so on -- all thermodynamic properties. The different shape of the volume, however, can result in a deviation $\Delta\area = \area^1 - \area^0 \neq 0$ for the surface area. This argument can be made completely rigorously in the case of a planar interface, supposing that we use Gibbs thermodynamics. For the spherical surface, it is much less convincing as it requires a sort of argument that is unusual in statistical mechanics~\cite{LFHMJ15}. But as long as we accept it, the Helmholtz free energy associated with each of the phases, $\phasea\helmholtz$ and $\phaseb\helmholtz$, will be the same for systems 0 and 1, while the interfacial excess contribution will be different
\begin{equation}
   \Delta\excess\helmholtz = \helmholtz^{\excesslabel, 1} - \helmholtz^{\excesslabel, 0} = (\helmholtz^1 - \phasea\helmholtz - \phaseb\helmholtz) - (\helmholtz^0 - \phasea\helmholtz - \phaseb\helmholtz) = \Delta\helmholtz.
\end{equation}
In the limit of infinitesimal perturbations, the micromechanical surface tension is obtained as
\begin{equation}
   \tensiondiff = \lim_{\Delta\area \to 0} \frac{\Delta\excess\helmholtz}{\Delta\area} = \lim_{\Delta\area \to 0} \frac{\Delta\helmholtz}{\Delta\area}.
\label{eqn:tensiondiff-micromechanical}
\end{equation}
This idea has also been expressed in the form of a partial derivative~\cite{SMMMJ10}
\begin{equation}
   \tensiondiff = \left(\frac{\partial\helmholtz}{\partial\area}\right)_{\absnumvec, \volume, \temperature};
\label{eqn:tensiondiff-nvt}
\end{equation}
this, again, can be formulated stringently in the planar case, where we may write $\helmholtz(\absnumvec, \volume, \temperature, \area)$ for the two-phase system, and the four arguments are independent thermodynamic degrees of freedom. In the spherical case, $\absnumvec$, $\volume$, and $\temperature$ determine $\area$, and the above to some extent becomes an abuse of notation -- it suggests a stronger basis of the method in thermodynamics than it actually has. In any case, whether we legitimately can write this or not, the definition from Eq.~(\ref{eqn:tensiondiff-micromechanical}) remains a well-defined micromechanical construct.

If this is implemented numerically directly in terms of the infinitesimal limit, it is most reasonable to call it a virial-based method. We speak of a test area method if this is implemented in the form of small finite perturbations, actually distorting the simulation volume by a small factor in different spatial directions. For example, Sampayo \etal~\cite{SMMMJ10} apply the test transformation
\begin{equation}
   \coordvec_\moli ~\mapsto~ \distortten\coordvec_\moli ~ = ~
      \begin{pmatrix}
         \sqrt{1 + \transf} & 0 & 0 \\ 0 & \sqrt{1 + \transf} & 0 \\ 0 & 0 & 1/[1+\transf] 
      \end{pmatrix}
         \coordvec_\moli
\label{eqn:test-transformation}
\end{equation}
to the spatial coordinates of the centre of mass for each molecule $\moli$. Momentum coordinates are not rescaled, so that $\Delta\energykin = 0$ and $\Delta\energyint = \Delta\energypot$; it is not made explicit what should be done with rotational quaternions or internal degrees of freedom, if present, but it seems reasonable enough that they should not be touched either. From the canonical partition function~\cite{SMMMJ10},
\begin{equation}
   \Delta\helmholtz ~=~ \avg{\Delta\energyint} ~-~ \frac{1}{2\temperature}\left(\avg{\Delta\energyint^2} - \avg{\Delta\energyint}^2\right) ~+~ \orderof(\transf^3),
\label{eqn:delta-helmholtz-1-2}
\end{equation}
where the first term is also denoted by $\Delta\helmholtz_1$, the second term by $\Delta\helmholtz_2$, and anything higher-order is being neglected. This method was primarily developed for planar interfaces~\cite{GJBM05}. However, Sampayo \etal~\cite{SMMMJ10} and Lau \etal~\cite{LFHMJ15} use it to compute the surface tension of spherical interfaces. The test transformation from Eq.~(\ref{eqn:test-transformation}) is taken to distort the sphere into an ellipsoid, and~\cite{LFHMJ15}
\begin{equation}
   \Delta\area = \frac{8\pi}{5}\radius^2\transf^2 + \orderof(\transf^3)
\label{eqn:test-area}
\end{equation}
is created as additional surface area,
so that for $\transf \to 0$, we can also write
\begin{equation}
   \left(\frac{\partial\area}{\partial[\transf^2]}\right)_{\absnumvec, \volume\temperature} = \frac{8\pi\radius^2}{5}.
\label{eqn:partial-area-transfsquare}
\end{equation}
As for the virial route, there are a few problems with this approach if the resulting surface tension $\tensiondiff$ is to be given a thermodynamic meaning. Moreoever, in the form presented above, it is not yet well-defined, since we can freely choose a definition of the radius to be used in Eqs.~(\ref{eqn:test-area}) and~(\ref{eqn:partial-area-transfsquare}); in this way, we can produce any result that we like for $\tensiondiff$. However, for a \textit{given} notion of the radius, it does provide a method and definition for $\tensiondiff$ that can be accepted as \textit{micromechanical}, \ie, as a microscopic observable that is not necessarily assigned a thermodynamic interpretation.

In the literature, when specifying $\Delta\area$ following Eq.~(\ref{eqn:test-area}), the equimolar radius $\radius = \radiusequim$ is used~\cite{SMMMJ10, LFHMJ15}. The results disagree with those from the virial route to such an extent that fixing the discrepancy just through the definition of the radius would lead to using an unreasonable dividing surface~\cite{SMMMJ10}. So, for our purposes, the main immediate aim is to figure out what makes the two methods disagree so much, and what else can be attempted in order to make them agree. 

\subsection{The virial route including the second-order virial}
\label{subsec:second-order-virial}

An observation from Sampayo~\etal~\cite{SMMMJ10}, which they consider a key insight, is that for droplets, in Eq.~(\ref{eqn:delta-helmholtz-1-2}), both $\Delta\helmholtz_1 = \avg{\Delta\energyint}$ and $\Delta\helmholtz_2 = -\mathrm{Var} \, \Delta\energyint/2\temperature$ deliver a major contribution to the micromechanical surface tension: They are both comparably large quantities with similar magnitude and opposite signs, from which $\tensiondiff$ of the curved interface emerges as a small difference. (Higher-order terms beyond $\Delta\helmholtz_2$ remain negligible.) This stands in contrast to the test area method for the planar interface, where $\Delta\helmholtz_1$ determines $\tensiondiff$, and all other terms, including $\Delta\helmholtz_2$, are much smaller in magnitude~\cite{SMMMJ10}. So there is a second-order term that plays a significant role in a spherical geometry, where its analogue in a planar geometry is much less significant.

Sampayo~\etal~\cite{SMMMJ10} also draw attention to the fact that when applying different distortions of the type in Eq.~(\ref{eqn:test-transformation}), transforming the sphere into an ellipsoid, the surface area always increases, so that the area as a function of the transformation parameter $\transf$ goes through a minimum for $\transf = 0$, \ie, in the case of no transformation. Again, this is not the case for a planar geometry, where transformation in one direction will increase the surface area, while an opposite transformation will decrease it. Consequently, as Lau \etal~\cite{LFHMJ15} point out, for small values of $\transf$, the surface area produced by distorting the sphere is approximately proportional to the square transformation parameter $\Delta\area \sim \transf^2$, \cf~Eq.~(\ref{eqn:test-area}); as opposed to the case of a planar interface normal to the $\coordz$ axis, where from Eq.~(\ref{eqn:test-transformation}), $\Delta\area \sim \transf$ by construction due to the way the test transformation is specified. Only for spherical interfaces, thre is a second-order contribution.

In the planar setting, the virial route expression for the surface tension is strictly obtained from the test area formalism in the infinitesimal $\transf \to 0$ limit. Naturally, the two methods then consequently produce the same results. We can take the same limit for the spherical test area method; it will give us a virial expression that, by design, must agree with the test area method. Based on the above, we can expect it to contain a second-order expression that was absent for planar interfaces, and that is indeed the case. We will call it the second-order virial tensor $\virialten_2$, as opposed to the first-order virial tensor $\virialten = \virialten_1$. Where $\virialten_1$ is a second-order tensor, $\virialten_2$ is a fourth-order tensor, and we can imagine these as being part of a series over $\natiter\in\naturals$, defined by
\begin{equation}
   \virialten_\natiter(\coordvec, \orientvec) ~=~ ~ \frac{(-1)^\natiter}{\natiter!} \sum_{\{\moli, \molj\}}
      \distancevec_{\moli\molj}^{\otimes\natiter} \otimes \nabla_{\distancevec_{\moli\molj}}^{\otimes\natiter} \, \potentialij(\distancevec_{\moli\molj}, \orientveci, \orientvecj),
\end{equation}
or, for the individual elements of the tensor,
\begin{equation}
   \virial_\natiter^{\axis_1\cdots\axis_{2\natiter}}(\coordvec, \orientvec) ~ = ~ \frac{(-1)^\natiter}{\natiter!} \sum_{\{\moli, \molj\}} 
      \distanceij^{\axis_1}\cdots\distanceij^{\axis_\natiter} \, \frac{\partial^\natiter\potentialij}{\partial\distanceij^{\axis_{\natiter+1}} \cdots \partial\distanceij^{\axis_{2\natiter}}}.
\label{eqn:kth-order-virial-tensor}
\end{equation}
Then, the scalar $\natiter$-th order virials are
\begin{equation}
   \virial_\natiter(\coordvec, \orientvec) = \virial_\natiter^{\coordx\cdots\coordx}(\coordvec, \orientvec) + \virial_\natiter^{\coordy\cdots\coordy}(\coordvec, \orientvec) + \virial_\natiter^{\coordz\cdots\coordz}(\coordvec, \orientvec).
\end{equation}
They all have the dimension of an energy. We proceed with Lau~\etal's expansion~\cite{LFHMJ15}
\begin{equation}
   \Delta\energyint(\coordvec, \orientvec) = \laua(\coordvec, \orientvec)\transf + \laub(\coordvec, \orientvec)\transf^2 + \orderof(\transf^3);
\label{eqn:lauexpansion}
\end{equation}
note that this is a microscopic observable. From spherical symmetry, $\avg{\laua} = 0$ (see below), and~\cite{LFHMJ15}
\begin{eqnarray}
   \Delta\helmholtz_1 & = & \avg{\Delta\energyint}
                      = \avg{\laub}\transf^2 + \orderof(\transf^3), \\
   \Delta\helmholtz_2 & = & - \, \frac{\mathrm{Var} \, \Delta\energyint}{2\temperature} 
                      = - \, \frac{\avg{\laua^2}}{2\temperature}\transf^2 + \orderof(\transf^3),
\end{eqnarray}
or, to summarize~\cite{LFHMJ15},
\begin{eqnarray}
   \left(\frac{\partial\helmholtz}{\partial[\transf^2]}\right)_{\absnumvec, \volume, \temperature}
      = \avg{\laub} - \frac{\avg{\laua^2}}{2\temperature},
\label{eqn:df-dchisquare}
\end{eqnarray}
whereas it is just $\avg{\laua}$ in the planar case~\cite{LFHMJ15}. The statistical mechanical observable for $\laua$ is
\begin{equation}
   \laua(\coordvec, \orientvec) = \sum_{\{\moli, \molj\}} \frac{\partial\potentialij}{\partial\transf}
      = - \sum_\axis \frac{\diff\distort^{\axis\axis}}{\diff\transf} \virial_1^{\axis\axis}.
\label{eqn:eval-laua}
\end{equation}
$\distort^{\coordx\coordx} = \distort^{\coordy\coordy} = \sqrt{1+\transf}$ and $\distort^{\coordz\coordz} = 1/(1+\transf)$ are the coefficients from the distortion matrix $\distortten$ in Eq.~(\ref{eqn:test-transformation}), the derivatives are taken at $\transf = 0$, and the index $\axis$ runs over $\{\coordx, \coordy, \coordz\}$. Consequently,
\begin{equation}
   \laua(\coordvec, \orientvec) = \virial_1^{\coordz\coordz} - \frac{\virial_1^{\coordx\coordx} + \virial_1^{\coordy\coordy}}{2},
\end{equation}
\ie, it is the same virial anisotropy expression that would directly yield $\Delta\helmholtz$ in the case of a planar interface oriented normally to the $\coordz$ axis. In a spherical geometry, all the three axes are equivalent, and the contributions will on average cancel out; hence, $\avg{\laua} = 0$.

For the variance of $\laua$, observe again that the axes are equivalent, so
\begin{eqnarray}
   \avg{\laua^2} & = & \frac{1}{2}\avg{
      \frac{1}{2}(\virial_1^{\coordx\coordx})^2 + \frac{1}{2}(\virial_1^{\coordy\coordy})^2 + 2(\virial_1^{\coordz\coordz})^2
         + \virial_1^{\coordx\coordx}\virial_1^{\coordy\coordy} - 2\virial_1^{\coordx\coordx}\virial_1^{\coordz\coordz}
            - 2\virial_1^{\coordy\coordy}\virial_1^{\coordz\coordz}
   } \nonumber\\
   & = & \frac{1}{2}\avg{
      (\virial_1^{\coordx\coordx})^2 + (\virial_1^{\coordy\coordy})^2 + (\virial_1^{\coordz\coordz})^2
         - \virial_1^{\coordx\coordx}\virial_1^{\coordy\coordy} - \virial_1^{\coordx\coordx}\virial_1^{\coordz\coordz}
            - \virial_1^{\coordy\coordy}\virial_1^{\coordz\coordz}
   } \nonumber\\
   & = & \frac{1}{2}\avg{\virial_1^2} - \frac{3}{2} \sum_{\{\axis, \axisb\}} \avg{\virial_1^{\axis\axis}\virial_1^{\axisb\axisb}};
\end{eqnarray}
for the last step, recall that $\virial_1(\coordvec, \orientvec)^2 = (\virial_1^{\coordx\coordx} + \virial_1^{\coordy\coordy} + \virial_1^{\coordz\coordz})^2$. Analogously, from Eq.~(\ref{eqn:lauexpansion}),
\begin{equation}
   \laub(\coordvec, \orientvec) = \frac{1}{2} \frac{\partial}{\partial\transf} \sum_{\{\moli, \molj\}} \frac{\partial\potential_{\moli\molj}}{\partial\transf}
      = - \, \frac{1}{2} \frac{\partial}{\partial\transf} \sum_\axis \frac{\diff\distort^{\axis\axis}}{\diff\transf} \virial_1^{\axis\axis}
         = - \, \frac{1}{2} \sum_\axis \frac{\diff^2\distort^{\axis\axis}}{\diff\transf^2} \virial_1^{\axis\axis}
            - \, \frac{1}{2} \sum_\axis \frac{\diff\distort^{\axis\axis}}{\diff\transf} \frac{\partial\virial_1^{\axis\axis}}{\partial\transf},
\end{equation}
compare Eq.~(\ref{eqn:eval-laua}). From the definition of the virial, we find $\laub(\coordvec, \orientvec) = \laub_1 + \laub_2 + \laub_3$ with
\begin{eqnarray}
   \laub_1(\coordvec, \orientvec) &=& - \, \frac{1}{2} \sum_\axis \frac{\diff^2\distort^{\axis\axis}}{\diff\transf^2} \virial_1^{\axis\axis}
                      ~=~ \frac{1}{8}\virial_1^{\coordx\coordx} + \frac{1}{8}\virial_1^{\coordy\coordy} - \virial_1^{\coordz\coordz}, \\
   \laub_2(\coordvec, \orientvec) &=& \, \frac{1}{2} \sum_\axis \frac{\diff\distort^{\axis\axis}}{\diff\transf} \sum_{\{\moli, \molj\}}
                             \left(\frac{\partial}{\partial\transf} \distort^{\axis\axis}\distanceij^{\axis, \transf=0}\right) \frac{\diff\potentialij}{\diff\distanceij}
                      ~=~ - \frac{1}{8}\virial_1^{\coordx\coordx} - \frac{1}{8}\virial_1^{\coordy\coordy} - \frac{1}{2}\virial_1^{\coordz\coordz}, \\
   \laub_3(\coordvec, \orientvec) &=& \, \frac{1}{2} \sum_\axis \frac{\diff\distort^{\axis\axis}}{\diff\transf} \sum_{\{\moli, \molj\}}
                            \distanceij^{\axis, \transf=0} \frac{\partial}{\partial\transf} \frac{\diff\potentialij}{\diff\distanceij}
                      \,=\, - \frac{1}{2} \sum_\axis \frac{\diff\distort^{\axis\axis}}{\diff\transf} \sum_{\{\moli, \molj\}}
                            \distanceij^{\axis, \transf=0} \sum_\axisb
                               \frac{\partial\distanceij^\axisb}{\partial\transf}\frac{\partial\forceij}{\partial\distanceij^\axisb}.
\end{eqnarray}
The third term, $\laub_3(\coordvec, \orientvec)$, contains a second derivative of the pair potential. This contribution has been overlooked by all virial-route methods for spheres so far, since through forces acting between molecules $\moli$ and $\molj$, with $\forceij^\axis = -\partial\potentialij/\partial\distanceij^\axis$, they were only accounting for first derivatives of $\potentialij$. To be consistent with the test area method, however, we need to include a term that contains the derivative of the force. This finds its straightforward expression in terms of $\virialten_2$
\begin{eqnarray}
   \laub_3(\coordvec, \orientvec) & = & \sum_\axis \sum_\axisb \frac{\partial\distort^{\axis\axis}}{\partial\transf}
      \frac{\partial\distort^{\axisb\axisb}}{\partial\transf} \virial_2^{\axis\axisb\axis\axisb} \nonumber \\
      & = & \frac{1}{4} \virial_2^{\coordx\coordx\coordx\coordx} + \frac{1}{4} \virial_2^{\coordy\coordy\coordy\coordy} + \virial_2^{\coordz\coordz\coordz\coordz}
         + \frac{1}{2} \virial_2^{\coordx\coordy\coordx\coordy} - \virial_2^{\coordx\coordz\coordx\coordz} - \virial_2^{\coordy\coordz\coordy\coordz},
\end{eqnarray}
where we use the symmetry relation $\virial_2^{\axis\axisb\axis\axisb} = \virial_2^{\axisb\axis\axisb\axis}$ that holds for the second-order virial tensor by construction, \cf~Eq.~(\ref{eqn:kth-order-virial-tensor}). For the ensemble averages, using the equivalence of axes,
\begin{eqnarray}
   \avg{\laub_1} & = & - \, \frac{1}{4} \avg{\virial_1}, \\
   \avg{\laub_2} & = & - \, \frac{1}{4} \avg{\virial_1}, \\
   \avg{\laub_3} & = & \frac{1}{2} \avg{\virial_2} - \frac{1}{2} \sum_{\{\axis, \axisb\}} \avg{\virial_2^{\axis\axisb\axis\axisb}},
\end{eqnarray}
which adds up to
\begin{equation}
   \avg{\laub} = - \, \frac{1}{2} \avg{\virial_1} + \frac{1}{2} \avg{\virial_2} - \frac{1}{2} \sum_{\{\axis, \axisb\}} \avg{\virial_2^{\axis\axisb\axis\axisb}}.
\end{equation}
The observable that needs to be sampled in a virial-route computation of the surface tension is
\begin{equation}
   \avg{\laub - \frac{\laua^2}{2\temperature}} = \frac{1}{2} \avg{
      \virial_2 - \virial_1 - \frac{\virial_1^2}{2\temperature} + \sum_{\{\axis, \axisb\}} \left(
         \frac{3\virial_1^{\axis\axis}\virial_1^{\axisb\axisb}}{2\temperature} - \virial_2^{\axis\axisb\axis\axisb}
      \right)
   }.
\end{equation}
Hence, only a single scalar quantity needs to be sampled. Then, with Eqs.~(\ref{eqn:tensiondiff-nvt}), (\ref{eqn:partial-area-transfsquare}), and (\ref{eqn:df-dchisquare}),
\begin{equation}
   \tensiondiff = \left(\frac{\partial\helmholtz}{\partial[\transf^2]} \, \frac{\partial[\transf^2]}{\partial\area}\right)_{\absnumvec, \volume, \temperature}
      = \frac{5\avg{\laub - \laua^2/2\temperature}}{8\pi\radius^2}.
\end{equation}
This is in addition to whatever approach is used to determine $\radius$. Section~\ref{sec:phenomenological} discusses this issue.

\section{Thermodynamic basis of the statistical mechanical expressions}
\label{sec:phenomenological}

In this section, we will be considering different ways in which the surface tension of the spherical interface can be defined within the framework of Gibbs thermodynamics. We will begin by reproducing the approach introduced previously~\cite{Horsch25}, which is essentially Kondo's approach~\cite{Kondo55, Kondo56} in a notation more suitable for the present discussion, and develop it in the direction needed for discussing the validity of methods like the spherical test area method (Section~\ref{subsec:spherical-test-area}), the second-order virial route (Section~\ref{subsec:second-order-virial}), and any other methods that compute $\tensiondiff$ as a free energy derivative for a system consisting of a dispersed phase nucleus and a surrounding bulk phase.

\subsection{Dividing surface notion and interfacial excess properties}
\label{subsec:dividing-surface-notion}

We now consider the case of a single spherical nucleus of a dispersed phase $\dispersedlabel$ at equilibrium with another fluid phase $\bulklabel$ that surrounds it. We can call $\bulklabel$ the surrounding phase or the bulk phase, even though both names are problematic in what they suggest: The phase $\bulklabel$ surrounds $\dispersedlabel¼$, but in our analysis still belongs to the system, not the surroundings; neither is it necessarily large in volume. In Gibbs thermodynamics, the interface is a strictly two-dimensional volume: While it can have excesses in other extensive quantities, it does not have any excess volume, so that $\volume = \dispersed\volume + \bulk\volume$. However, its position as such is not specified by the physics of the system -- instead, the same thermodynamic formalism can be applied to different notions of the interface.

By the notion $\notion$ we understand a function that, given knowledge of the microscopic physics (such as the density profile, \etc), determines a radius $\radius_\notion$ of the dispersed phase, and consequently the surface area $\area_\notion = 4\pi\radius_\notion^2$, the volume of the dispersed phase $\dispersed\volume_\notion = 4\pi\radius_\notion^3/3$, the volume of the surrounding phase $\bulk\volume_\notion = \volume - \dispersed\volume_\notion$, and many more. With very few exceptions, all interfacial properties depend on $\notion$, which means that they are not uniquely specified by the thermodynamic state of the system, but depend on what we select as the position of the two-dimensional effective dividing surface. This is in principle a long-established approach, introduced by Kondo~\cite{Kondo55, Kondo56} in the 1950s. To choose $\notion$ means to impose one additional, non-physical boundary condition that, within reason, can be chosen freely to simplify the thermodynamic analysis. 

Suppose thermodynamic equilibrium between the two phases within a system subject to suitable boundary conditions consistent with a stable phase equilibrium, such as canonical boundary conditions. The intensive quantities that are the same for the two phases are the temperature $\temperature$ and the chemical potential $\chempotvec$ (denoted by a vector because there can be multiple components). We are not assuming here that $\chempotvec$ and $\temperature$ are externally imposed boundary conditions; nonetheless, they can be used as descriptors of the thermodynamic state: All intensive properties of the coexisting phases can be given as a function of $\chempotvec$ and $\temperature$. If one extensive property of one of the phases is known, for example its volume, then together with $\chempotvec$ and $\temperature$ its extensive properties are determined as well. The \textit{interfacial excess} of the extensive property $\extensive$ is then defined by
\begin{equation}
   \excess\extensive_\notion = \extensive - \dispersed\extensive_\notion - \bulk\extensive_\notion.
\label{eqn:excess}
\end{equation}
There, $\extensive$ is the property of the whole system. It is defined by the thermodynamic state and does not depend on $\notion$, as opposed to all other quantities in the equation that we can, within reason, set to any value we like by choosing $\notion$ accordingly. (Except for $\extensive = \volume$, where $\excess\volume = 0$.) Note that while the extensive properties depend on $\notion$, the intensive properties do not, since they depend on $\chempotvec$ and $\temperature$ only. Hence, in $\dispersed\extensive_\notion = \dispersed\density\dispersed\intensive\dispersed\volume_\notion$, where $\intensive$ is the molar quantity corresponding to $\extensive$, only $\dispersed\volume_\notion$ depends on the choice of dividing surface; and the same for phase $\bulklabel$.

The two most important dividing surface notions are defined in terms of interfacial excess properties: For the equimolar surface ($\notion$ = $\equimolarlabel$), the excess amount of substance is zero
\begin{equation}
   \sum_\component \excess\absnum_{\component, \equimolarlabel} = 0,
\end{equation}
where $\component$ runs over components. The Laplace surface ($\notion = \laplacelabel$) is defined by the Laplace equation
\begin{equation}
   \laplace\radius = \left(\frac{\grandpot^\excesslabel_\laplacelabel}{2\pi[\dispersed\pressure(\chempotvec, \temperature) - \bulk\pressure(\chempotvec, \temperature)]}\right)^{1/3}.
\label{eqn:laplace-excess}
\end{equation}
$\grandpot$ is the grand potential, and $\grandpot^\excesslabel_\laplacelabel$ is its interfacial excess for the Laplace surface.
The surface tension can be defined in different ways, three of which are developed below: The absolute surface tension $\tensionabs_\notion$; the total-differential surface tension $\tensiondiff_\notion$; the partial-differential surface tension $\tensionpart^\boundary_\notion$.

\subsection{Absolute and total-differential surface tension}

The \textit{absolute surface tension} $\tensionabs_\notion$ can be defined in terms of the grand potential
\begin{equation}
   \tensionabs_\notion = \frac{\grandpot^\excesslabel_\notion}{\area_\notion},
\label{eqn:tensionabs}
\end{equation}
as its excess per surface area. With this, the Laplace equation can be given in its usual form
\begin{equation}
   \Delta\pressure = \frac{2\laplace\tensionabs}{\laplace\radius},
\label{eqn:laplace-absolute}
\end{equation}
for the Laplace radius $\laplace\radius$, \cf~Eq.~(\ref{eqn:laplace-excess}), where $\Delta\pressure = \dispersed\pressure - \bulk\pressure$. Since the grand potential of a homogeneous system is $\grandpot = -\pressure\volume$, we can rewrite Eq.~(\ref{eqn:excess}) with Eq.~(\ref{eqn:tensionabs}) as
\begin{equation}
   \grandpot = - \dispersed\pressure\dispersed\volume_\notion - \bulk\pressure\bulk\volume_\notion + \tensionabs_\notion\area_\notion.
\label{eqn:grandpot-tensionabs}
\end{equation}
The \textit{total-differential surface tension} $\tensiondiff_\notion$ is defined by the differential
\begin{equation}
   \diff\helmholtz = \chempotvec\,\diffvec\absnumvec - \dispersed\pressure\,\diff\dispersed\volume_\notion - \bulk\pressure\,\diff\bulk\volume_\notion - \entropy\,\diff\temperature + \tensiondiff_\notion\,\diff\area_\notion.
\label{eqn:tensiondiff}
\end{equation}
Above, $\dispersed\volume_\notion$ and $\area_\notion$ are the same thermodynamic degree of freedom,
$\diff\dispersed\volume_\notion = \radius_\notion\,\diff\area_\notion/2$, so that
\begin{equation}
   \diff\helmholtz = \chempotvec\,\diffvec\absnumvec - \bulk\pressure\,\diff\bulk\volume_\notion - \entropy\,\diff\temperature + \left(\tensiondiff_\notion - \frac{\radius_\notion\dispersed\pressure}{2}\right) \diff\area_\notion,
\end{equation}
and we might attempt to construct an expression similar to Eq.~(\ref{eqn:tensiondiff-nvt}), such as $\tensiondiff_\notion = \dispersed\pressure\radius_\notion/2 + \partial\helmholtz/\partial\area_\notion$ at constant $\absnumvec$, $\bulk\volume_\notion$, and $\temperature$, in order to determine $\tensiondiff_\notion$. However, with $\compnum$ components, that would require $\compnum + 3$ independent variables, namely, $\absnumvec$, $\bulk\volume_\notion$, $\temperature$, and $\area_\notion$. In reality, counting both extensive and intensive properties, there are only $\compnum + 2$ independent thermodynamic degrees of freedom; \eg, we can vary $\absnumvec$, $\volume$, and $\temperature$ independently, and they together will control all properties of the system and both phases. Hence, we cannot simply obtain $\tensiondiff_\notion$ as a partial derivative of Helmholtz free energy; the construction from Eq.~(\ref{eqn:tensiondiff-nvt}) fails for the same reason.

However, we can extend the argument as follows to include non-equilibrium states: Under canonical boundary conditions, for given $\absnumvec$, $\volume$, and $\temperature$, and for a given notion of the dividing surface, the two phases $\dispersedlabel$ and $\bulklabel$ can partition the volume in different ways. Knowing that the equilibrium shape is spherical, we will here allow only spherical geometries. The equilibrium state is the one for which $\helmholtz$ is minimal. With $\diff\bulk\volume_\notion = \diff\volume - \diff\dispersed\volume_\notion$, continuing the operation from above, the total differential of Helmholtz free energy in $\absnumvec$, $\volume$, $\temperature$ and $\area_\notion$ becomes
\begin{equation}
   \diff\helmholtz = \chempotvec\,\diffvec\absnumvec - \bulk\pressure\,\diff\volume - \entropy\,\diff\temperature + \left(\tensiondiff_\notion - \frac{\radius_\notion\Delta\pressure}{2}\right) \diff\area_\notion.
\label{eqn:helmholtz-four}
\end{equation}
Since $\absnumvec$, $\volume$, and $\temperature$ are imposed as constant boundary conditions, the equilibrium state corresponding to a minimum in $\helmholtz$ requires the coefficient $\tensiondiff_\notion - \radius_\notion\Delta\pressure/2$ to be zero, \ie,
\begin{equation}
   \Delta\pressure = \frac{2\tensiondiff_\notion}{\radius_\notion}.
\label{eqn:laplace-tensiondiff}
\end{equation}
This proves that for the surface tension $\tensiondiff_\notion$ defined from Eq.~(\ref{eqn:tensiondiff}), the Laplace equation holds for \textit{any} dividing surface, as opposed to $\tensionabs_\notion$, where it only holds for the Laplace surface ($\notion = \laplacelabel$).

While this is in itself a desirable and unsurprising result, it also poses two serious problems. First, the test area method (Section~\ref{subsec:spherical-test-area}) seems to be developed for a total-differential definition of the surface tension. And if we include non-equilibrium configurations, we can indeed take a partial derivative of the form given by Sampayo~\etal~\cite{SMMMJ10}, \cf~Eq.~(\ref{eqn:tensiondiff-nvt}). However, it becomes
\begin{equation}
   \left(\frac{\partial\helmholtz}{\partial\area_\notion}\right)_{\absnumvec, \volume, \temperature} ~=~ \tensiondiff_\notion - \frac{\radius_\notion\Delta\pressure}{2} ~=~ 0;
\end{equation}
Sampayo~\etal~\cite{SMMMJ10} explicitly require this to be $\tensiondiff$, not zero.

Second, \eg~in a canonical-ensemble simulation, we can determine $\chempotvec$ from methods that do not require any definition of a dividing surface. Both $\chempotvec$ and $\temperature$ are then known, and the intensive properties of $\dispersedlabel$ and $\bulklabel$, including $\dispersed\pressure$ and $\bulk\pressure$, can be determined from these through an equation of state. So the choice of dividing surface $\notion$ does not affect $\Delta\pressure$. However, this means that for different $\notion$, the total-differential surface tension is simply proportional to the radius of tension
\begin{equation}
   \frac{\tensiondiff_\notion}{\laplace\tensiondiff} = \frac{\radius_\notion}{\laplace\radius}.
\label{eqn:tensiondiff-proportionality}
\end{equation}
This is naturally not in line with what the test area method does either. There, the radius is not clearly defined; the equimolar radius is used arbitrarily in the absence of such a definition. However, if a smaller radius is chosen, it will result in a greater surface tension, and vice versa.

So while the test area method appears to be constructed to compute the surface tension from Eq.~(\ref{eqn:tensiondiff}), that is not what it does. This makes it hard to assign any thermodynamic meaning to the $\tensiondiff$ from that method. This consequently applies to the new method from Section~\ref{subsec:second-order-virial} as well.

\subsection{Gibbs adsorption equation and nucleation theorem}

We can differentiate Eq.~(\ref{eqn:grandpot-tensionabs}), Legendre-transform Eq.~(\ref{eqn:tensiondiff}) from $\helmholtz$ to $\grandpot$, and then subtract
\begin{eqnarray}
   \diff\grandpot & = & - \dispersed\pressure\,\diff\dispersed\volume_\notion - \dispersed\volume_\notion\,\diff\dispersed\pressure - \bulk\pressure\,\diff\bulk\volume_\notion - \bulk\volume_\notion\diff\bulk\pressure + \tensionabs_\notion\,\diff\area_\notion + \area_\notion\,\diff\tensionabs_\notion, \\
   \diff\grandpot & = & - \absnumvec\,\diffvec\chempotvec - \dispersed\pressure\,\diff\dispersed\volume_\notion - \bulk\pressure\,\diff\bulk\volume_\notion - \entropy\,\diff\temperature + \tensiondiff_\notion\,\diff\area_\notion, \\
   0 & = & \absnumvec\,\diffvec\chempotvec - \dispersed\volume_\notion\,\diff\dispersed\pressure - \bulk\volume_\notion\diff\bulk\pressure + \entropy\,\diff\temperature + (\tensionabs_\notion - \tensiondiff_\notion) \diff\area_\notion + \area_\notion\,\diff\tensionabs_\notion.
\label{eqn:gibbs-ads-intermediate}
\end{eqnarray}
Now consider Eq.~(\ref{eqn:gibbs-ads-intermediate}) for each of the two phases separately, \ie, the Gibbs-Duhem equation
\begin{eqnarray}
   0 & = & \dispersed\absnumvec_\notion\,\diffvec\chempotvec - \dispersed\volume_\notion\,\diff\dispersed\pressure + \dispersed\entropy_\notion\,\diff\temperature, \\
   0 & = & \bulk\absnumvec_\notion\,\diffvec\chempotvec - \bulk\volume_\notion\,\diff\bulk\pressure + \bulk\entropy_\notion\,\diff\temperature,
\end{eqnarray}
and subtract these two from Eq.~(\ref{eqn:gibbs-ads-intermediate}) to obtain a generalized Gibbs adsorption equation
\begin{eqnarray}
   \excess\absnumvec_\notion\,\diffvec\chempotvec + \excess\entropy_\notion\,\diff\temperature + (\tensionabs_\notion - \tensiondiff_\notion) \diff\area_\notion + \area_\notion\,\diff\tensionabs_\notion = 0
\end{eqnarray}
or, in terms of the adsorption $\adsorptionvec_\notion = \excess\absnumvec_\notion/\area_\notion$ and the surface entropy $\surfentrop_\notion = \excess\entropy_\notion/\area_\notion$,
\begin{equation}
   \diff\tensionabs_\notion ~=~ - \adsorptionvec_\notion\,\diffvec\chempotvec - \surfentrop_\notion\,\diff\temperature + \frac{2(\tensiondiff_\notion - \tensionabs_\notion)}{\radius_\notion} \, \diff\radius_\notion,
\label{eqn:generalized-gibbs-adsorption}
\end{equation}
and for the Laplace surface ($\laplace\tensiondiff = \laplace\tensionabs$)
\begin{equation}
   \diff\laplace\tensiondiff ~=~ - \laplace\adsorptionvec\,\diffvec\chempotvec - \laplace\surfentrop\,\diff\temperature.
\end{equation}
This is the \textit{Gibbs adsorption equation} in its common form, from which at constant temperature, the Tolman equation can be developed~\cite{Tolman49b}. Similarly, a generalized form of the Tolman equation~\cite{Horsch25} can be developed from the generalized Gibbs adsorption equation as given in Eq.~(\ref{eqn:generalized-gibbs-adsorption}).

The \textit{nucleation theorem}~\cite{Kashchiev82, OK94}, on the other hand, does not explicitly involve any interfacial excess quantities -- it is consequently valid for any arbitrary notion of the dividing surface. Instead of \textit{interfacial excess} quantities, \cf~Eq.~(\ref{eqn:excess}), it is formulated in terms of \textit{nucleus formation} quantities, which for any extensive property $\extensive$ and associated molar property $\intensive$ are defined as
\begin{equation}
   \nucform\extensive = \extensive - \bulk\density\bulk\intensive\volume;
\label{eqn:nucform}
\end{equation}
\ie, the deviation between the actual value of $\extensive$ for the system containing both the dispersed and surrounding phases, on the one hand, and a reference state where the surrounding phase takes up the whole volume, on the other hand. The nucleation theorem relates $\nucform\grandpot$ to $\nucform\absnumvec$ (where the latter is often approximated by the critical nucleus size, $\nucform\absnumvec \approx \dispersed\absnumvec$, in the case of vapour-liquid nucleation). For the grand potential of nucleus formation, Eq.~(\ref{eqn:nucform}) becomes
\begin{eqnarray}
   \nucform\grandpot & = & \grandpot + \bulk\pressure\volume, \\
   \diff\nucform\grandpot & = & \diff\grandpot + \bulk\pressure\,\diff\volume + \volume\,\diff\bulk\pressure.
\end{eqnarray}
With Eqs.~(\ref{eqn:helmholtz-four}) and (\ref{eqn:laplace-tensiondiff}), and inserting $\diff\bulk\pressure = \bulk\densityvec\,\chempotvec$,
\begin{eqnarray}
   \diff\grandpot & = & - \absnumvec\,\diffvec\chempotvec - \bulk\pressure\,\diff\volume - \entropy\,\diff\temperature, \\
   \diff\nucform\grandpot & = & (\bulk\densityvec\volume - \absnumvec) \diffvec\chempotvec - \entropy\,\diff\temperature,
\end{eqnarray}
wherein $\bulk\densityvec\volume - \absnumvec = -\nucform\absnumvec$, resulting for an isothermal transition in the nucleation theorem
\begin{equation}
    \diff\nucform\grandpot \,=\, -\nucform\absnumvec\,\diffvec\chempotvec.
\end{equation}

\subsection{Kondo's equation}

The nucleus formation quantities are directly related to interfacial excess quantities
\begin{equation}
   \nucform\extensive = \excess\extensive_\notion + (\dispersed\density\dispersed\intensive - \bulk\density\bulk\intensive)\dispersed\volume\notion,
\label{eqn:nucform-excess}
\end{equation}
\begin{figure}[b!]
    \centering
    \includegraphics[width=0.9333\textwidth]{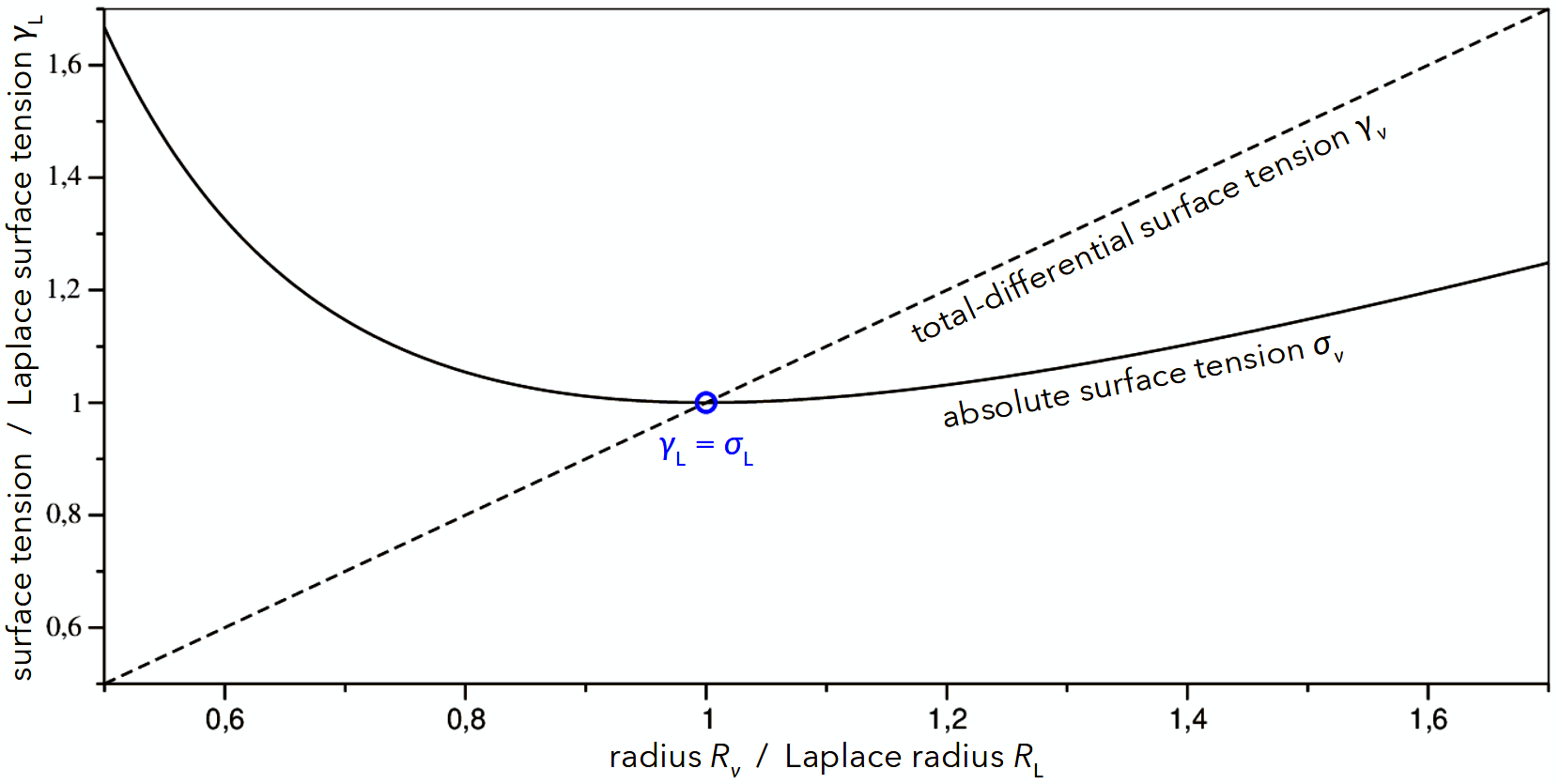}
    \caption{Absolute and total-differential surface tensions as a function of the notion of the dividing surface, from Eqs.~(\ref{eqn:tensiondiff-proportionality}) and~(\ref{eqn:kondo}); compare also Fig.~2 from Kondo~\cite{Kondo56}.}
    \label{fig:gamma-nu_r-nu_plot}
\end{figure}
which follows from the definitions of these quantities by Eqs.~(\ref{eqn:excess}) and (\ref{eqn:nucform}). To apply this to $\nucform\grandpot$ and $\excess\grandpot_\laplacelabel$ for the grand potential, we first define an extended form of the Laplace equation
\begin{equation}
   \Delta\pressure = \frac{2\tensionabs_\notion}{\radius_\notion} + \pressurediffabsnu,
\label{eqn:laplace-extended}
\end{equation}
wherein $\pressurediffabsnu = 2(\tensiondiff_\notion - \tensionabs_\notion)/\radius_\notion$. With this, using $\area_\notion = \excess\grandpot_\notion/\tensionabs_\notion = 3\dispersed\volume_\notion/\radius_\notion$, Eq.~(\ref{eqn:nucform-excess}) becomes
\begin{eqnarray}
   \nucform\grandpot & = & \excess\grandpot_\notion - \Delta\pressure\dispersed\volume_\notion \label{eqn:nucform-grandpot-excess} \\
      & = & \frac{1}{3}\excess\grandpot_\notion - \pressurediffabsnu\dispersed\volume_\notion.
\label{eqn:nucform-grandpot-solution}
\end{eqnarray}
For the Laplace surface, $\nucform\grandpot = \excess\grandpot_\laplacelabel/3$, a well-known relation from classical nucleation theory. Now subtract one third of Eq.~(\ref{eqn:nucform-grandpot-excess}) from Eq.~(\ref{eqn:nucform-grandpot-solution}):
\begin{equation}
   \frac{2}{3}\nucform\grandpot = \left(\frac{\Delta\pressure}{3} - \pressurediffabsnu\right)\dispersed\volume_\notion.
\end{equation}
Resolving this for $\pressurediffabsnu$ and inserting into Eq.~(\ref{eqn:laplace-extended}) produces
\begin{equation}
   \Delta\pressure = \frac{2\tensionabs_\notion}{\radius_\notion} + \left(\frac{\Delta\pressure}{3} - \frac{2\nucform\grandpot}{3\dispersed\volume_\notion}\right),
\end{equation}
which for $\tensionabs_\notion$ resolves to
\begin{equation}
   \tensionabs_\notion = \frac{\radius_\notion\Delta\pressure}{3} + \frac{\nucform\grandpot}{4\pi\radius^2_\notion}.
\label{eqn:pre-kondo}
\end{equation}
Here we can insert the relations that hold for the Laplace surface: The Laplace equation, $\Delta\pressure = 2\tensionabs_\laplacelabel/\radius_\laplacelabel$, and $\nucform\grandpot = \excess\grandpot_\laplacelabel/3 = 4\pi\radius^2_\laplacelabel\tensionabs_\laplacelabel/3$. We obtain Kondo's equation~\cite[therein, Eq.~3.23]{Kondo56}
\begin{equation}
  \frac{\tensionabs_\notion}{\laplace\tensionabs} = \frac{2}{3} \frac{\radius_\notion}{\laplace\radius} + \frac{1}{3} \left(\frac{\radius_\notion}{\laplace\radius}\right)^{-2},
\label{eqn:kondo}
\end{equation}
which describes how the absolute surface tension varies as a function of $\notion$ at constant thermodynamic conditions. Fig.~\ref{fig:gamma-nu_r-nu_plot} visualizes Eq.~(\ref{eqn:tensiondiff-proportionality}) together with Eq.~(\ref{eqn:kondo}); from this and all of the above, there can be no doubt what quantity we must be interested in obtaining: The surface tension for the Laplace dividing surface, where the absolute and total-differential definitions of the surface tension become equivalent, $\laplace\tensionabs = \laplace\tensiondiff$. This is commonly understood, but it deserves pointing out that the above analysis only confirms it. Any method that computes a surface tension other than this must either permit relating it to $\laplace\tensiondiff$, so that it can be determined indirectly, or it is from the standpoint of interfacial \textit{thermodynamics} completely useless and can only play a role within a different formalism that can make proper use of the respective \textit{micromechanics}.

\section{Conclusion: Open questions}
\label{sec:conclusion}

It is much easier to devise, implement, and run methods that would compute a free-energy derivative, such as by a virial or test transformation approach, than to compute an interfacial excess free energy. To compute the free energy difference between the multi-phase system the homogeneous systems, a method needs to facilitate thermodynamic integration between the homogeneous and heterogeneous states; this is costly, as the whole transition needs to be sampled. Binder's method does this by grand-canonical sampling~\cite{BDOVB10}, and thereby determines $\excess\grandpot$.

So a derivative-based method is in principle desirable. The open question from this can be posed in two ways. First, it is not clear how we should express the Laplace surface tension as a free energy derivative. Since $\laplace\tensionabs = \laplace\tensiondiff$, this could be done from any of the expressions for $\tensionabs_\notion$ or for $\tensiondiff_\notion$, but the development from Section~\ref{sec:phenomenological} does not provide us with an obvious starting point. Instead, we found that Eq.~(\ref{eqn:tensiondiff-nvt}) does \textit{not} work within the system of variables developed here -- we obtained a zero where we should have obtained $\tensiondiff$ according to Sampayo~\etal~\cite{SMMMJ10}. Second, we already have a derivative-based method, namely, the test-area method, and now in addition to this, the foundations for a second-order virial route that is in agreement with the test-area method if the reasoning from Section~\ref{subsec:second-order-virial} is correct. The test-area method is validated empirically; its results agree with those from Binder's method, and are consistent with the thermodynamic analysis of density profiles in terms of the excess equimolar radius as developed in previous work~\cite{HHSAEVMJ12, HH14a} (as opposed to results from the standard, first-order virial route~\cite{VKFH06, HVH08}, which are therefore \textit{known} to be wrong). But a proper theoretical grounding is missing.

The total-differential surface tension cannot be directly obtained as a partial derivative of a thermodynamic potential; at least, from the discussion above, it is not clear how this can be consistently done. The same is the case for the absolute surface tension. Therefore, it may make sense to account for a formalism where the surface tension \textit{can} be obtained in that way, namely,
\begin{equation}
   \tensionpart^\boundary_\notion ~=~ \left(\frac{\partial\excess\grandpot_\notion}{\partial\area_\notion}\right)_\boundary,
\label{eqn:tensionpart}
\end{equation}
where $\boundary$ is a collection of variables for which we can write $\excess\grandpot_\notion(\area_\notion, \boundary)$ such that the arguments can all be independently varied while maintaining thermodynamic equilibrium, and such that they jointly uniquely define $\excess\grandpot_\notion$.
We can refer to this as the \textit{partial-differential surface tension}.

This formalism will require case distinctions concerning what variables are dependent or independent, which depends on the choice of $\notion$. For example, with $\boundary = \{\chempotvec, \temperature\}$, the partial-differential surface tension would be ill-defined at the Laplace surface, since there, the surface area can be given as a function $\laplace\area(\chempotvec, \temperature)$; so it is not possible to take the derivative from Eq.~(\ref{eqn:tensionpart}). On the other hand, for other $\notion$, the same expression is well-defined; in these cases, Eq.~(\ref{eqn:tensionpart}) can be shown to reduce to the Laplace equation, so that
\begin{equation}
   \tensionpart^{\{\chempotvec, \temperature\}}_\notion = \tensiondiff_\notion, \quad \forall\notion \in \{\notion \,\mid\, \area_\notion, \chempotvec, \temperature \,\, \textnormal{are independent}\}.
\end{equation}
So it is in fact possible to obtain $\tensiondiff_\notion$ as a free energy derivative for some $\notion$, just not for $\notion = \laplacelabel$. Together with some closure relation, it might be possible to connect $\radius_\notion$ to $\laplace\radius$, and $\tensiondiff_\notion$ to $\laplace\tensiondiff$.

Or should we call on a demon to save us from our predicament? It might just work. Newton, the alchemist, popularized the idea of external entities intervening into bodies' equations of motion when he argued that once planets threaten to leave their foreseen orbits, God ever so slightly pushes them back. But as we know well, one's God is another's demon. That other has been Lord Maxwell -- though Maxwell called it a \textit{finite being}, and it was Lord Kelvin who first called it a \textit{demon}. After Maxwell's and Szil\'ard's demon~\cite{MNV09}, McDonald's demon~\cite{McDonald63}, Toda's demon~\cite{HVR95}, and many more, we would here be dealing with Sampayo's demon. The demon would distort the system in a way similar to the action of an external field, so that the ellipsoidal shape is the \textit{equilibrium} shape. This would introduce an additional variable so that, \eg, $\transf$ could become a \textit{thermodynamic} boundary condition. It is plausible that the theory for this scenario can be developed from an analogy between the demon's action and an electric field.

\bigskip

\noindent
\textit{Acknowledgment.\/} CECAM is acknowledged for funding the CECAM Workshop on Interfacial Properties: Open Questions (IPOQ 2025) at UKRI STFC Daresbury Laboratory, Cheshire, UK.

{\footnotesize
   \bibliographystyle{klunummod}
   \bibliography{ipoq-2025-notes-virial}
}

\end{document}